\documentclass[preprint]{aastex}

\begin{document}

\title{M31 Globular Clusters in the HST Archive: I. Cluster Detection and Completeness%
\footnote{Based on observations made with 
the NASA/ESA Hubble Space Telescope,
obtained from the data archive at Space Telescope Science Institute. 
STScI is operated by the Association of Universities
for Research in Astronomy, Inc. under NASA contract NAS 5-26555.}
}

\author{Pauline Barmby\altaffilmark{2}
\ \& John P. Huchra}
\affil{Harvard-Smithsonian Center for Astrophysics, 60 Garden St., Cambridge, MA 02138} 
\email{pbarmby@cfa.harvard.edu, huchra@cfa.harvard.edu}
\altaffiltext{2}{Guest User, Canadian Astronomy Data Centre, 
which is operated by the Herzberg Institute of Astrophysics, National Research Council of Canada.}

\shortauthors{Barmby \& Huchra}
\shorttitle{M31 globulars in the HST Archive I}

\begin{abstract}

Globular clusters at the distance of M31 have
apparent angular sizes of a few arcseconds.  
While many M31 GCs have been detected and studied from ground-based images,
the high spatial resolution of HST allows much more robust
detection and characterization of star cluster properties.
We present the results of a search of 157 HST/WFPC2 images of M31,
We found 82 previously-cataloged globular cluster candidates as well as 32 new
globular cluster candidates and 20 open cluster candidates.
We present images of the new candidates and photometry
for all clusters. We assess existing cluster catalogs' completeness
and use the results to estimate the total number of GCs in M31
as $460\pm70$. The specific frequency is 
$S_N=1.2\pm0.2$ and the mass specific frequency $T=2.4\pm0.4$;
these values are at the upper end of the range seen for spiral galaxies.

\end{abstract}


\section{Introduction}

Globular clusters (GCs) are among the oldest surviving
stellar objects in the universe. They provide collections
of Population~II stars with homogeneous abundances 
and histories, and unique stellar dynamical conditions.
The Milky Way's globular cluster system (GCS) is the
prototypical one, and its study has contributed much to 
our knowledge of stellar evolution and galactic structure.
It is important to make sure that conclusions drawn from
this study are not biased
either because the Milky Way's GCS is somehow unusual 
or because our location in the Galaxy prevents us from fully 
characterizing its properties.
Globular clusters in Local Group galaxies are particularly
valuable for comparison with Milky Way globular clusters.
M31 has the Local Group's largest
globular cluster population, so it is a natural starting place for
studies of extragalactic globular clusters.

The first M31 globular cluster catalog was published by \citet{hub32},
followed by \citet{sn45}, \citet{vet62a}, \citet{sar77},
and \citet{cra85}. The most comprehensive recent catalog is that of
\citet{bat87}; recent works by \citet{bat93} and \citet{m98} cover only
portions of M31. All of these catalogs contain objects which are not 
M31 globular clusters: for example, Table~2 of \citet{b00} lists
199 cluster candidates later shown to be non-clusters. The existing catalogs
are also likely to be missing clusters due to magnitude, spatial coverage, 
and/or resolution limits. \citet{bat93} defined several samples of M31
globular clusters, including a `confirmed' sample (199 objects),
an `adopted best' sample (298 objects), and an `extended' sample
(356 objects). In \citet{b00} we compiled a list of clusters and plausible
candidates containing 435 objects.

Quantifying the extent of incompleteness and contamination
in M31 globular cluster catalogs is extremely important for 
the interpretation of globular cluster system properties.
For example, the spatial distribution of known clusters is flatter,
and their globular cluster luminosity function
(GCLF) brighter, near the nucleus \citep{bat93,bhb01} --- 
is this because the clusters there are truly fewer and brighter,
or because existing surveys have not detected the entire cluster
population? Even the census of Milky Way clusters is likely to
be incomplete: \citet{m95} estimates that 10--30 Milky Way globulars may be
hidden behind the Galactic bulge 
and therefore missing from current catalogs, which list about 150
objects \citep{h96}. Two such clusters were found by \citet{hur00}.
It is not unreasonable to suspect that the M31 cluster catalogs could
be incomplete by at least a similar fraction.

Ground-based high-resolution imaging and spectroscopy have been used to distinguish M31 
globular clusters from interlopers such as foreground stars, background 
galaxies, and other objects belonging to M31 (e.g., H~II regions and open clusters). 
The bright ($V\lesssim17$) portion of M31 globular cluster catalogs has been fairly 
thoroughly examined using one or both of these methods.
\citet{rac91} and \citet{rac92} used short-exposure CCD images taken in
excellent seeing to determine if cluster candidates in the M31 halo
were resolved into stars; they found that majority of the halo cluster
candidates were background galaxies, not clusters.
Radial velocities from optical spectroscopy have also been used by several 
groups \citep[e.g.,][]{hss82,fmf90,hbk91,fed93,b00} to eliminate background galaxies
and foreground stars from cluster candidate lists. 
Neither method is infallible, however: compact clusters 
may be mistaken for background galaxies if not resolved into stars, or for stars
if they have a small radial velocity.% 
\footnote{Recall that M31 has a heliocentric
radial velocity $v_r\approx-300$~km~s$^{-1}$. The velocity range of M31 globular clusters
is about +70 to $-700$~km~s$^{-1}$, and the Galactic models of 
\citet*{rcb89} predict that the radial velocities of Milky Way 
stars with similar colors and magnitudes to M31 globulars are 
in the range $-400$ to $+100$ km~s$^{-1}$.}
HST imaging, with its superior spatial resolution, is a useful
tool for removing some of the ambiguities inherent in the
ground-based studies.

At the distance modulus of M31 given by \citet{sg98} and \citet{hol98}
($(m-M)_0=24.47$, $d=783$~kpc), the angular resolution of HST's
WFPC2 camera is equivalent to a spatial resolution of 0.38~pc. 
is very helpful for the identification of globular clusters in M31. 
The differences between globular clusters and contaminating objects
are much more obvious than with ground-based imaging. 
M31 has been a popular target for HST: as of 1 December 2000, 
the Hubble Data Archive contained almost 1100 WFPC2 images
within 150\arcmin\ of the center of the galaxy.
As of the same date, about two dozen M31 globular clusters had been
specifically targeted for observation with HST, and the images
of these clusters comprise about 20\% of all the M31 images.
The goal of most targeted HST observations of 
M31 globular clusters has been the production of color-magnitude 
diagrams for the clusters and surrounding stellar populations.
HST programs which specifically targeted M31 globular clusters
include GOs 5112, 5420, 5464, 5907, 6477, 6671, 7826, 8296, and 8664.
Our study uses the publicly-available archival data from these programs
and many others.

In mid-2000, we began a project to search for globular clusters in archival
HST images for the purpose of quantifying the incompleteness
of existing cluster catalogs; preliminary results were
described in \citet{bhb01}. The present paper 
report the results of our efforts to find globular and other
star clusters in archival HST/WFPC2 images and their
implications for catalog completeness and contamination.
A companion paper \citep*{bh01} presents measurements of the 
structural parameters of the
clusters and their implications. We do not attempt to construct CMDs
for the clusters, since this work is already being carried out 
by other groups.

\section{Searching the HST archive}

\begin{figure}
\includegraphics[scale=0.7]{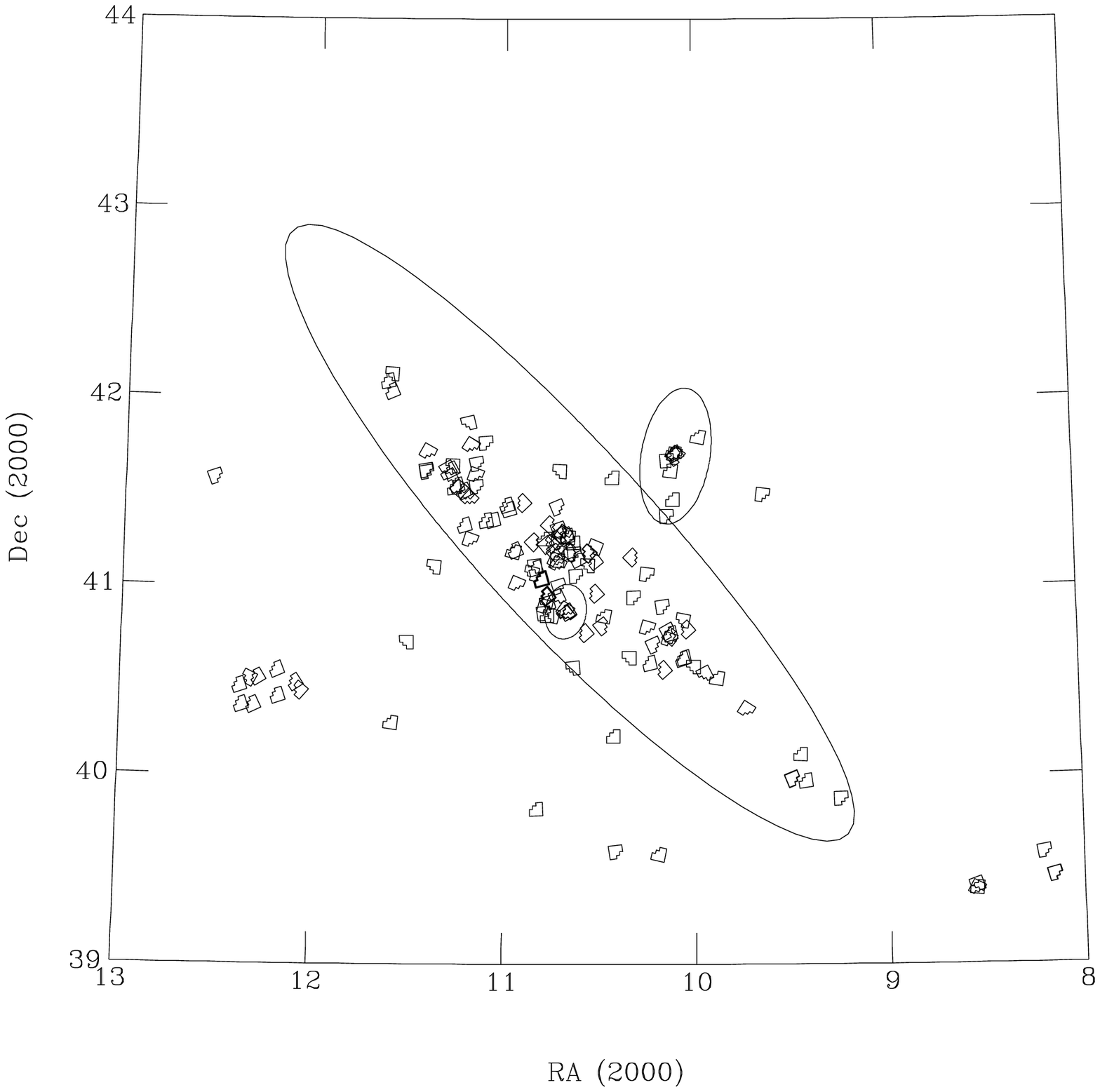}
\caption
{Location and orientation of M31 HST fields.
Large ellipse is M31 disk/halo boundary as defined by Racine (1991); smaller
ellipses are $D_{25}$ isophotes of NGC~205 (NW) and M32 (SE).
The WFPC2 symbols are drawn about
$1.5\times$ actual size to make them easier to see. The group of 
fields at $\alpha\sim12\arcdeg$, $\delta\sim40.5\arcdeg$ is part of a snapshot
survey of field galaxies (GO-6354).
\label{field-pic}}
\end{figure}

We searched the HST Archive for all WPFC2 observations with the
following properties:
\begin{itemize}
\item center of field within $<150$\arcmin\ of the center of M31 
\item broadband filter with central wavelength 300~nm or longer
\item total exposure time longer than 100~s.
\end{itemize}
These parameters were chosen to ensure that we would have a reasonable
chance of detecting globular clusters if they were in the image fields.
Many images met the requirements, but since
most positions had more than one observation per filter
and observations in more than one filter, the images comprised
only 157 separate fields. Some of these fields were known to contain
M31 globular clusters; we retained these fields in our
search as a check on our ability to identify clusters.
We searched the images in only one filter per field. 
If more than one filter was available, we chose filters 
in the following order: F555W, F814W, F606W, F450W, F439W, F336W, F300W. 
(This ordering reflects the distribution filters used for the images
combined with our desire to examine as many fields as possible in the same filter.)
Information on the fields searched, including dataset name, location,
filter, and exposure time is given in Table~\ref{tbl-flds}.
The images searched are mostly in F555W and F814W, although there
is at least one image in each of the filters listed above.
The exposure times ranged from 100 to 8400~s.
Figure~\ref{field-pic} shows the location of all fields
on the sky.

We retrieved the images from either the Space Telescope 
Science Institute or the Canadian Astronomy Data Centre.
In both cases the images were pipeline-processed from the raw
data at the time of retrieval with the best available
calibration images.
From STScI we retrieved individual HST images; when multiple images 
existed for a single field (e.g., in the case of `cosmic ray-split' images),
we combined the images using the IRAF task {\sc crrej}.
From CADC we retrieved `WFPC2 associations'; these are coadded
images produced by the CADC pipeline, 
which combines multiple CR-split images with the {\sc gcombine}
task. We found that {\sc gcombine} did not adequately remove
cosmic-ray hits when only two images were co-added, so in that
case we retrieved the individual images and combined them with {\sc crrej}.
There were no obvious differences in the images produced using
the two methods --- we used both since we became aware of 
availability of association images from the CADC part way into the project.

Once the images were processed, we began the search for star clusters.
The first step was carried out `blind', that is, without any knowledge
of the positions of cataloged clusters.
Working independently, each of us visually examined each image.
PB used SExtractor \citep{ba96} to automatically identify objects with 
large areas and/or extended profiles, then visually checked the SExtractor 
candidates (many of these were actually bright stars) and searched for
additional candidates. JH used only visual examination of the images. 
Bright M31 globular clusters can be visually distinguished from stars and
elliptical galaxies because they appear more `ragged' at the edges
(being resolved into stars) and do not have the diffraction spikes seen 
around bright stars. Faint or small clusters are distinguished more
by their image shapes --- larger than the point spread function, and
less smooth than a galaxy --- than by resolution into stars.
Clusters are distinguished from H~II regions or nebulae by the fact that the
latter are much more diffuse are show few individual stars.
The visual classification is not completely objective, but it
was the best method we could contrive for dealing with the
large number of images to be examined and the large number of potential
contaminating objects.

Our confidence in the visual classification was bolstered by the fact that
we only disagreed on the classification of about 10\% of the objects.
We re-examined these together to make a final classification.
We combined our two lists of cluster candidates to make
a final list. Although we were interested primarily in globular clusters, we recorded
positions of possible open clusters as well.
Following previous authors \citep[e.g.,][]{bat87,m98}, we classified our globulars
in classes A through D, where A is `very likely to be a globular cluster'
and D is `likely not a globular cluster'. We refer to objects in classes
A and B as  good candidates, and objects in classes C and D as marginal
cluster candidates.
After generating our final list of cluster candidates, we checked the image
positions against existing catalogs of M31 globular clusters.
This allowed us to gauge our detection efficiency and locate
objects we would otherwise have missed.
The globular cluster list used was a `master list' of globular clusters and candidates, 
produced by combining the lists of \citet{sar77}, 
\citet{cra85}, \citet{bat87}, \citet{bat93}, and \citet{m98};
it includes all the objects listed in the \citet{b00} catalog, plus
additional low-probability candidates and non-clusters. 

\section{Search results}

\subsection{Globular clusters}

We consider the low and high-probability globular clusters separately.
`High probability' are clusters A or B class  clusters
from \citet{bat87}, \citet{bat93}, or \citet{m98}; all other objects are
`low probability'. \citet{rac91} showed that 
the \citet{bat87} classification correlates well with the probability 
that a candidate will be subsequently shown to be a cluster.
75 high-probability clusters from our master list were located in 
the HST fields; we detected 71, and some images of previously-cataloged 
clusters are shown in Figure~\ref{fig-gc}. Three of the four non-detections
(138--000, 166--000, and 133--191) appeared to be stars or blends of stars 
rather than globular clusters; the fourth object was DAO040 and we
did not detect any object at the coordinates given by \citet{cra85}.
Of the 72 low-probability (class C or D) cluster candidates in our HST fields,
we found 7 good candidates (000--D038, 000--M91, 020D--089, 097D--000,
132-000, 264--NB19, and NB39), 4 marginal candidates 
(000--M045, 257--000, NB41 and NB86), and 45 objects which did not
appear to be clusters. We did not detect the other 14 objects
in our visual search. On re-examining the positions of these objects, 
we found that none were good or even marginal cluster
candidates. Several were clearly stars, and the others were blends of stars
or blank fields. Table~\ref{tbl-noncl} gives a list of the non-clusters
and their classifications.

\subsection{Uncataloged globular clusters}

Our visual search of the HST fields produced 32 objects not included in
any cluster catalog. 10 of these were good candidates, although only
about half are as obviously clusters as most of the brighter objects.
The good candidates' images are shown in Figure~\ref{fig-newclust}. 
The nature of the remaining 22 objects
is unclear. They are clearly not stars; all are at least marginally resolved
(FWHM $\gtrsim0.2$\arcsec). However, most are quite faint, and they are not obviously
resolved into stars as is the case for most of the globular clusters. They may be blended
stars in M31, compact background galaxies, or compact star clusters. We show images
of these low-quality objects in Figure~\ref{fig-myst}.
Table~\ref{tbl-newgc} gives the location and quality of all the new
cluster candidates.

\subsection{Open clusters}

The dividing line between open and globular clusters is somewhat blurred,
even in the Milky Way. In their compilation of data on Milky Way globular clusters,
\citet{dm93} note that there are several globulars (BH~176, UKS~2)
which could instead be open clusters. In our search, we noted several 
concentrated objects which could be M31 open clusters. Their nature
is uncertain: they could also be low-concentration globulars, or just chance superpositions
of stars. Their images are shown in Figure~\ref{fig-oc}. We checked the cluster coordinates against
those given in \citeauthor{hod79}'s (\citeyear{hod79}) list of M31 open clusters.
The coordinates in that catalog have rather low precision (20\arcsec\ in both right
ascension and declination), so we searched for coordinate matches within an error circle of 
radius 30\arcsec. We found 5 matches and attempted to confirm these by comparing
the finding charts in \citet{hod82} to our images. The results were inconclusive:
either the objects were not clearly identified on the charts, or they were
located too close to the edge of the HST image to make a positive identification.
We note the possible matches in Table~\ref{tbl-newgc}.

\begin{figure}
\includegraphics[scale=0.7]{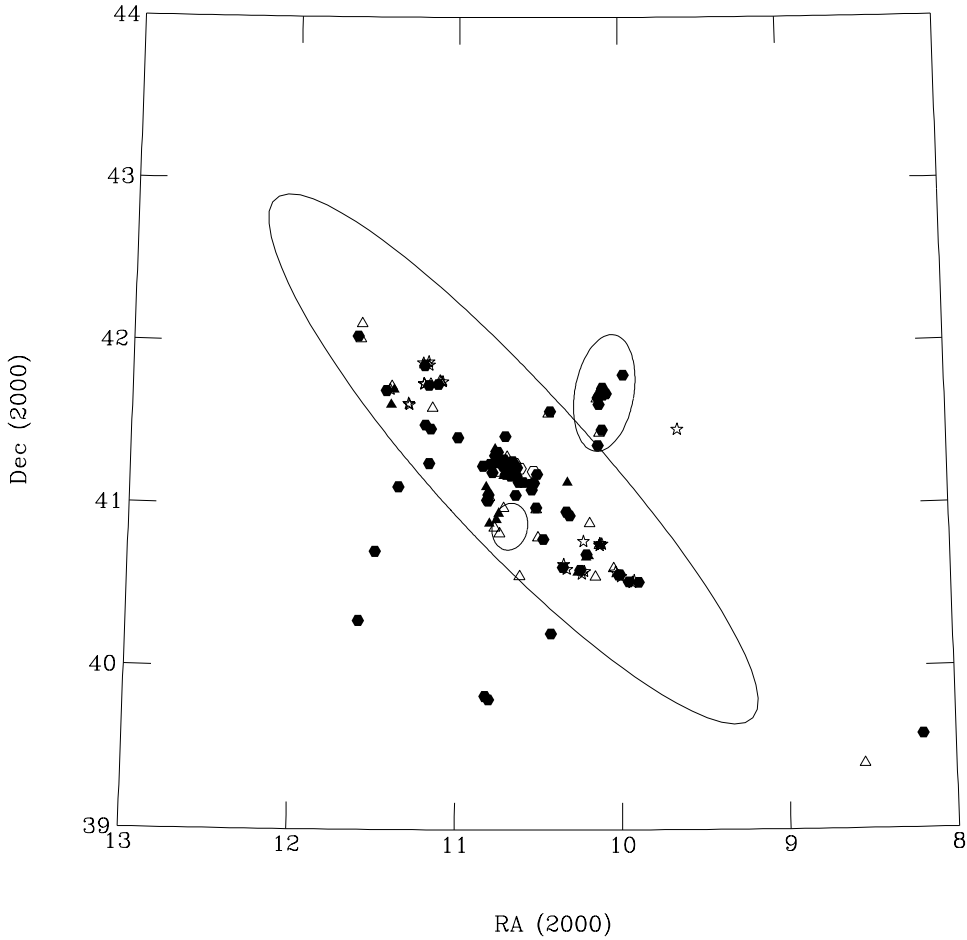}
\caption
{Position on the sky of all GCs, GC candidates and open clusters.
Ellipses are the same as in Figure~1. Filled symbols are good-quality GC candidates; 
open symbols are marginal candidates. Hexagons are previously-cataloged objects; 
triangles are newly-discovered objects; stars are possible open clusters.
\label{posplot}}
\end{figure}

To see if any of our newly-proposed globular and open cluster candidates had
been previously cataloged as background galaxies, we checked their positions 
against those of galaxies listed in NED.%
\footnote{The NASA/IPAC Extragalactic Database
(NED) is operated by the Jet Propulsion Laboratory, California
Institute of Technology, under contract with the National Aeronautics
and Space Administration} 
None of the new clusters matched the
position of any galaxy listed in NED, although one is listed as a possible
H~II region \citep{str92} and two others may contain radio and X--ray sources 
\citep{zha93,sup97}. The matches are also noted in Table~\ref{tbl-newgc}.  
However, the matches are uncertain since positional uncertainties for the other surveys 
are large. Figure~\ref{posplot} shows the positions on the sky of all the 
M31 clusters, both previously-known and newly-discovered. The `open cluster' 
near NGC~205 is well outside the disk and is probably not a real cluster.

\section{Integrated photometry\label{sec-intphot}}

After the M31 clusters had been identified on the `search' images, we retrieved
images of their fields in other available filters to
extract the most photometric information from the HST Archive.
All but 18 clusters had been imaged by WFPC2 in more than one filter.
We combined images for cosmic-ray rejection in the same manner used for
the search images. Additional processing steps included
removing cosmic rays interactively using the IRAF task {\sc imedit}
(this was especially important for non-cosmic-ray-split images) and
correcting for warm pixels using the IRAF task {\sc cosmicrays}.
While the {\sc stsdas} task {\sc warmpix} is the preferred method
of dealing with warm pixels, it is slow and requires correction
of individual images before they are combined for
cosmic ray rejection. Since we had 
hundreds of individual images to deal with, we chose the more expedient 
method of treating the warm pixels as if they were as cosmic ray hits
on ground-based images.
Nearby bright stars and CCD flaws were masked out of the images
to prevent contamination of the photometry.
A few images were not useful for photometry at all:
the globular clusters were either very faint (mostly in the F300W and F336W
filters), too close to an image edge, or saturated. 

Photometry of extragalactic globular clusters
is unfortunately not as simple as photometry of isolated stars or galaxies.
There are two key steps in integrated photometry of M31 clusters:
measuring the background light, and identifying an appropriate aperture size.
The background light consists of two components: unresolved light 
from the sky and M31, and light from resolved stars in M31
(the latter are a lesser problem in ground-based photometry of M31 GCs
since many fewer M31 stars are resolved). 
Standard background estimators are usually designed to determine the 
sky background level by rejecting the stars in the background annulus.
Since we expect there to be stars overlapping our clusters as well,
we estimated the background value for each image as the mean 
(rather than the more commonly
used median or mode) of the pixel values around the image edge, 
and subtracted it from the image before doing photometry.

\begin{figure}
\includegraphics[scale=0.7]{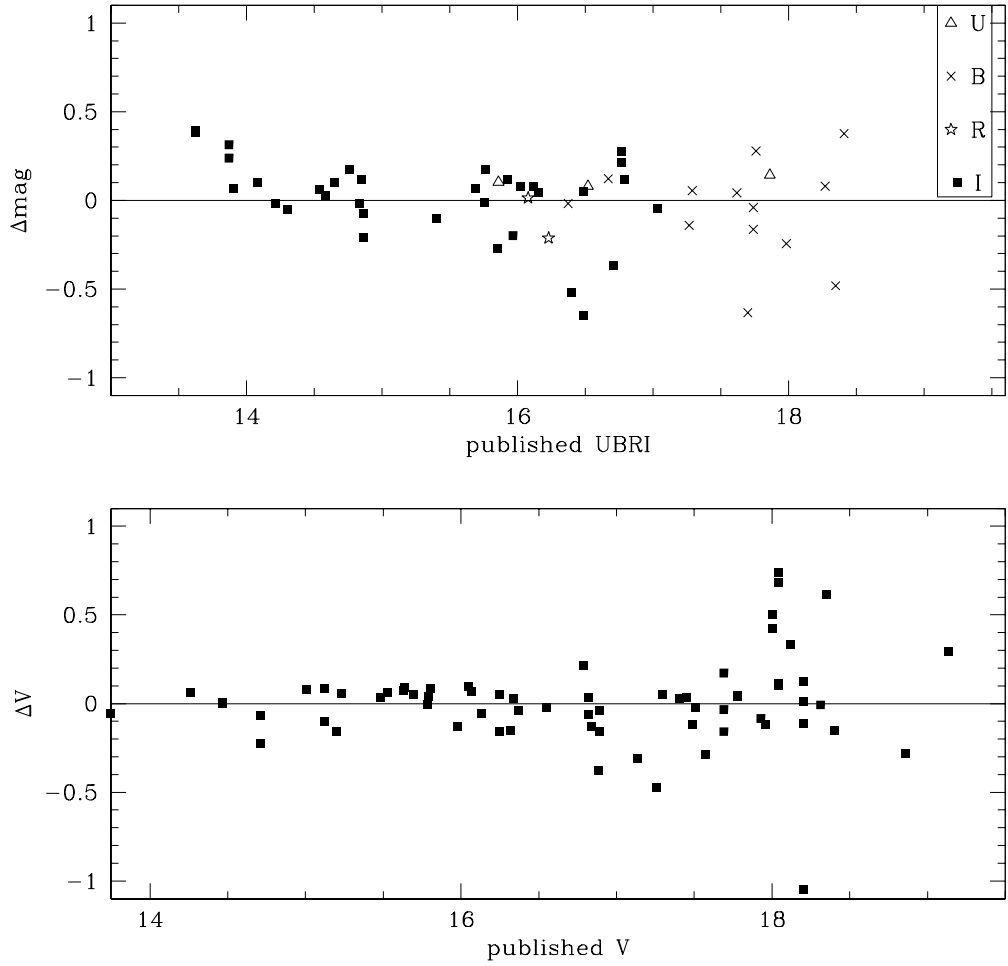}
\caption
{Comparison of integrated HST photometry to ground-based photometry:
vertical axis is (published photometry)--(HST photometry).
\label{fig-photcomp}}
\end{figure}

Determining the `correct' aperture size to be used for integrated
photometry is non-trivial since the clusters are not all the same size. 
We estimated the total flux for each object by measuring aperture magnitudes
in concentric apertures spaced 0.15\arcsec\ apart, plotting magnitude 
growth curves, and noting where the flux stopped increasing.
Using these measurements of the total flux of each cluster,
we determined the half-light radius%
\footnote{The half-light radius $r_h$
is that which contains half of the integrated cluster light. It 
should not be confused with the radius at which the surface brightness drops 
to half of its central value, called variously the core radius $r_c$, the 
half-intensity radius, or the half-width at half maximum (HWHM).}
by interpolating the aperture magnitude curves.
We calibrated the instrumental magnitudes from the WFPC2 system to
the standard system by iteratively solving the equations given in \citet{h95},
using the charge-transfer-efficiency corrections
given by \citet{dol00}. The iterative solution of the calibration equations
requires instrumental magnitudes in at least two filters; for objects with only one
instrumental magnitude, we fixed the `standard color' as either the measured
ground-based color from \citet{b00}, or if that was unavailable, 
the average M31 GC color.
The results for integrated magnitudes and half-light radii are given in 
Table~\ref{tbl-phot}. In Figure~\ref{fig-photcomp},
we compare the new HST photometry to the ground-based measurements
compiled in \citet{b00}.
The agreement is gratifying: the median offset in $V$ is 
$0.01\pm0.04$~magnitudes, and in $I$ is $0.06\pm0.04$~mag.
Most of the large offsets are for objects near the edge of a WFPC2 chip,
or whose previous photometry was estimated from photographic
plates.

\section{Completeness of globular cluster catalogs in M31\label{sec-comp}}

To estimate the completeness of globular cluster catalogs in M31, we first need 
to understand our own detection efficiency. We estimated this by inserting 
artificial globular clusters into the inner images, for which
the distance from the center of M31 $\equiv R_{gc}$ was less than 30\arcmin. 
The artificial clusters
were actually images of the brightest real globular clusters we detected.
To insert the artificial clusters, we scaled the image fluxes 0--4
magnitudes fainter, adjusted for the exposure time of the inserted image, rotated
the cluster to a random position angle, and applied a random axial ratio
from 0.85 to 1.0. This may not have been an entirely correct method of generating
artificial clusters, since cluster size, surface brightness, and integrated
magnitude are known to be correlated for Milky Way clusters. However, we decided
it was better not to introduce additional assumptions about the correlation
of these parameters into our detection test. 
Once an artificial cluster was inserted into a copy of each HST frame,
we extracted a $15\arcsec\times15$\arcsec\ region around the inserted cluster, 
and examined only that portion of the image. This cut-out procedure was similar
to the procedure used for the re-examination of `problem' images. In fact
the visual examination of the two groups of images was done at the same time,
with no reference to which were the inserted clusters and which were the 
real objects. For each cut-out image, we decided whether or not
it was a {\it bona fide} globular cluster.

\begin{figure}
\includegraphics[scale=0.7]{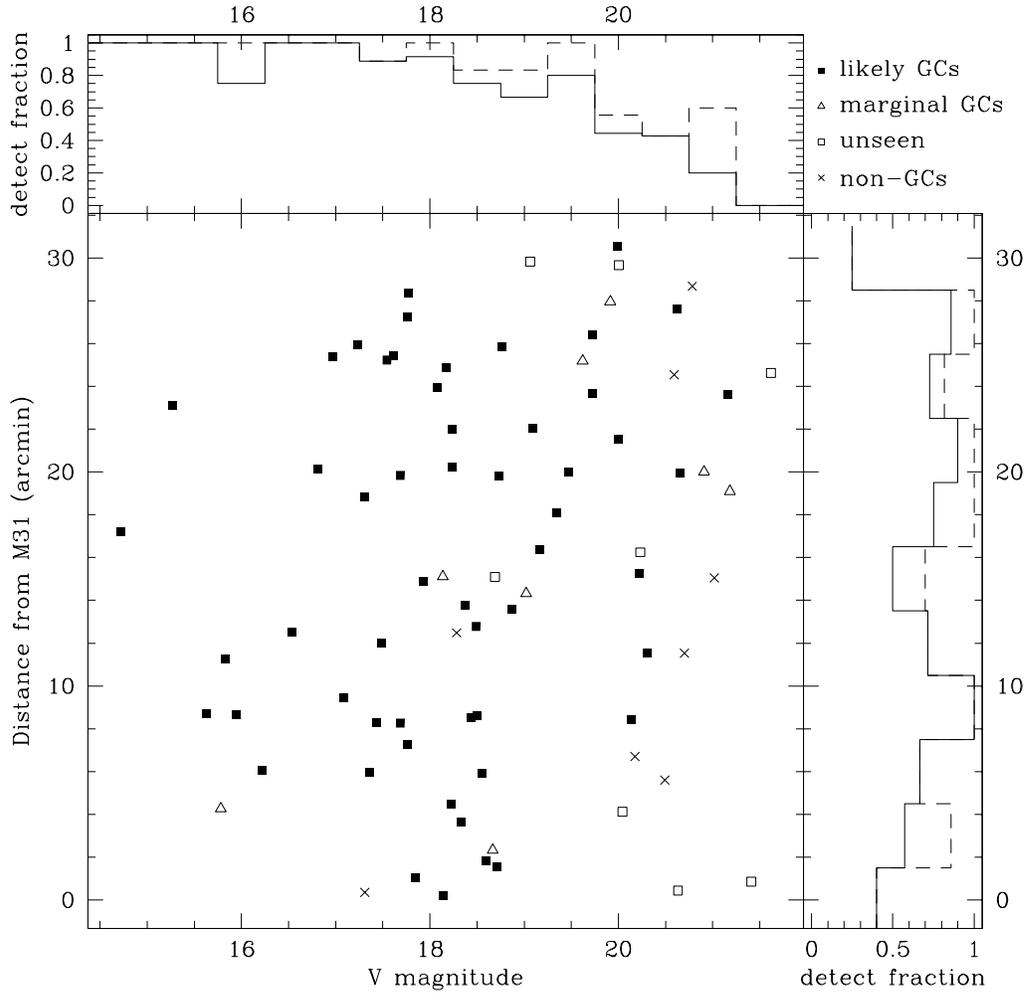}
\caption
{Measurement of globular cluster detection efficiency.
Large plot: $V$ magnitude vs. $R_{gc}$ for artificial clusters. Symbol
type indicates whether an object was detected and how it was classified.
The histograms are the fraction of inserted objects detected;
solid lines include only A or B class (`good') GCs, and dashed lines
include marginal objects.
\label{fig-incomp}}
\end{figure}

The results of our search for the inserted globular clusters  are
in Figure~\ref{fig-incomp}, where detection of each inserted cluster is
indicated as a function of $V$ magnitude and $R_{gc}$. 
The figure shows, as expected, that our detection 
efficiency was generally worse for fainter objects and objects near the
center of M31. Faint clusters are more difficult to find against the bright
background of the M31 disk and nucleus. We also failed to detect a few bright
objects, mostly in short exposures or in the near-UV filters
F300W and F336W. Overall, we correctly identified 80\% of the inserted 
clusters, and 92\% of the objects which appeared in long F555W and F814W exposures.

\begin{figure}
\includegraphics[scale=0.7]{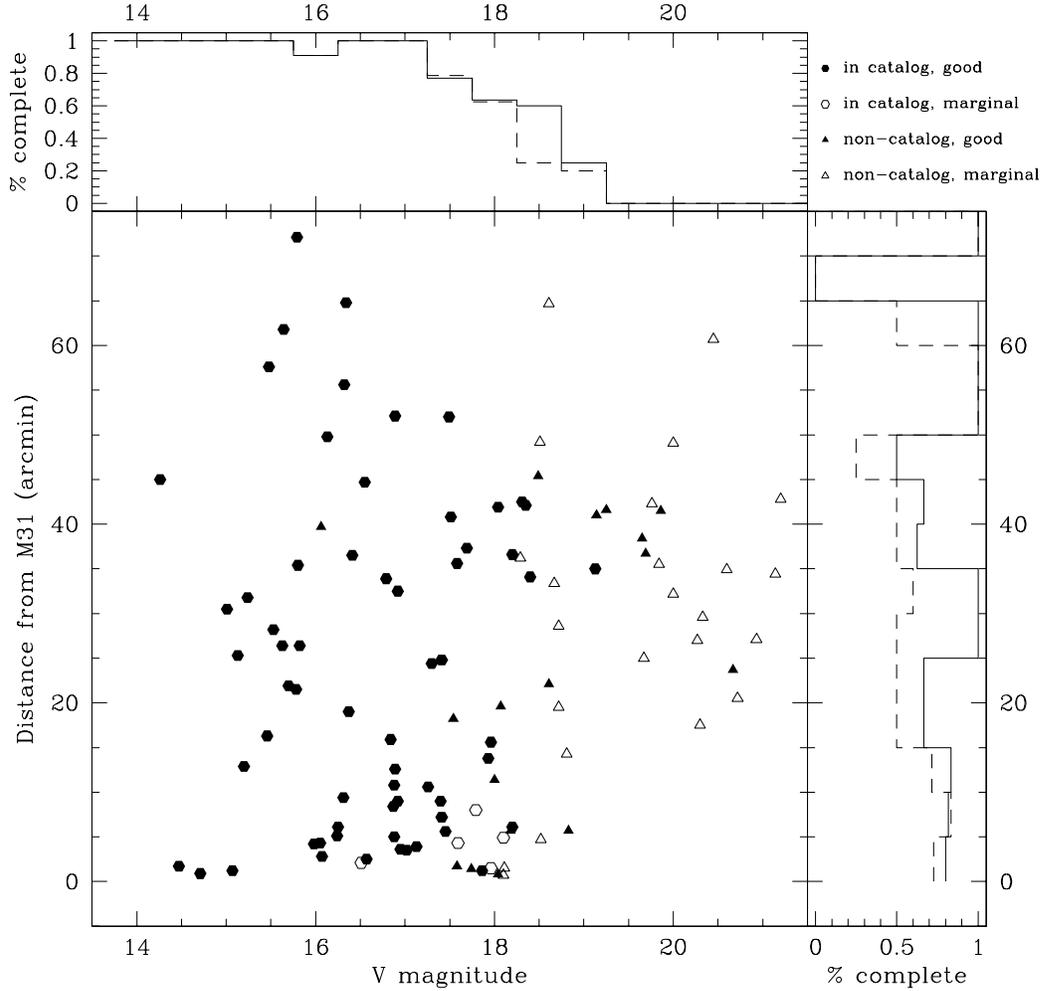}
\caption
{Location of previously-cataloged and newly-discovered M31
globular clusters in $V$ vs. $R_{gc}$ space.
Symbols indicate object quality and presence in existing catalogs.
Histograms estimate the existing catalogs' completeness by showing
(number of previously known objects per bin)/(number of known
$+$ number of new objects per bin).
Solid line histograms include only A or B class GCs, and dashed line
histograms include marginal objects.
\label{fig-gcdet}}
\end{figure}

\begin{figure}
\includegraphics[scale=0.7]{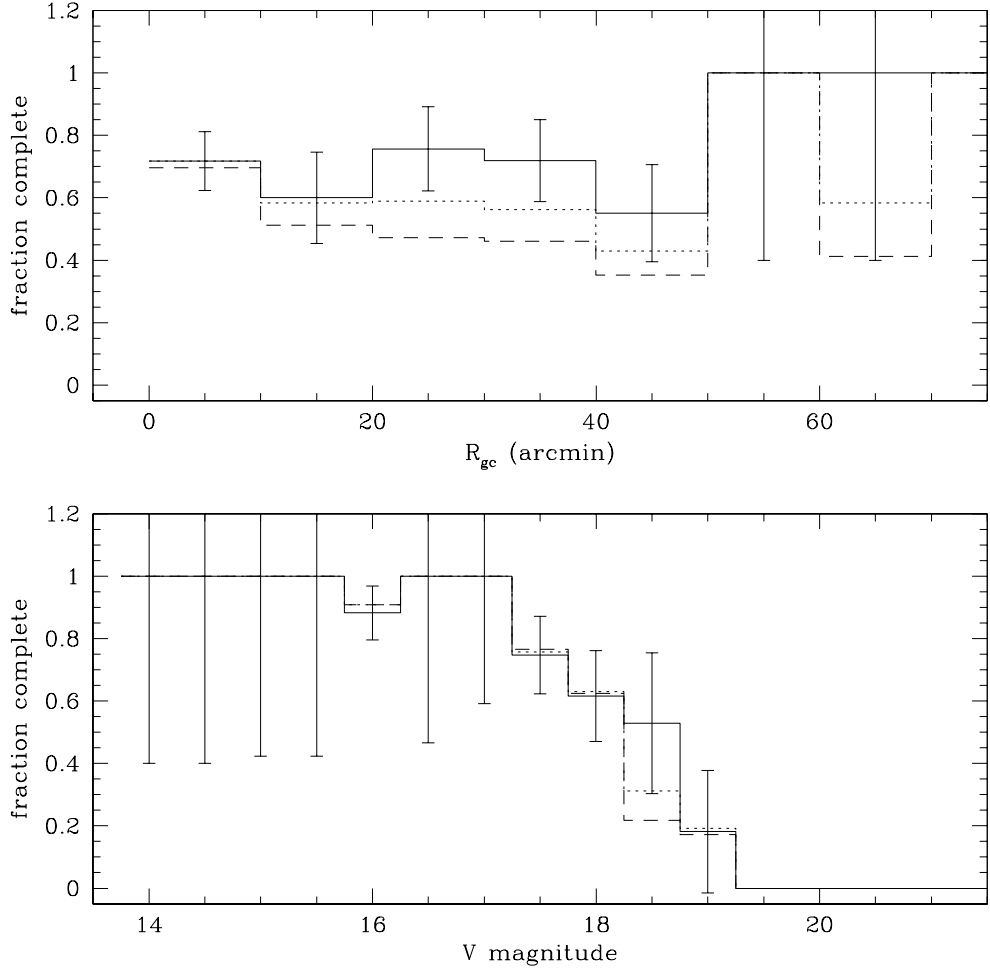}
\caption
{Completeness functions for existing surveys of globular clusters in M31.
Top panel, $C(R_{gc})$, is summed over entire magnitude range, and
bottom panel, $C(V)$, is summed over entire radial range.
Different line types reflect different assumptions about how many
marginal objects  are true M31 clusters. Solid lines: none, dotted lines: 
half, and dashed lines: all.
\label{fig-compfn}}
\end{figure}

The distribution of the real globular clusters and candidates 
detected in the HST images is shown in Figure~\ref{fig-gcdet}. 
The number of newly-detected objects increases at fainter magnitudes; there
is no clear trend in the number of new objects with $R_{gc}$.
We use the data in Figures~\ref{fig-incomp}~and~\ref{fig-gcdet} to estimate the completeness of
existing catalogs. While it would be desirable to estimate the completeness as a
joint function of magnitude and position, the small number of objects we have to work
with makes deriving $C(V,R_{gc})$ difficult. Instead we summed over one variable to produce
separate functions $C(V)$ and $C(R_{gc})$, which are plotted in Figure~\ref{fig-compfn}.
The catalog completeness is computed by dividing the number of cataloged objects in a given 
bin by the true number of objects: 
\begin{equation}
C=\frac{N_{\rm cat}}{N_{\rm true}}=\frac{N_{\rm cat}}{N_{\rm cat}+N_{\rm new}/\eta}
\end{equation}
where the `true' object total is the sum of the number of cataloged objects and the 
number of new objects divided by our detection efficiency $\eta$. 
The number of new objects includes the marginal objects. From the results
of \citet{rac91}, only a fraction ($f\lesssim0.5$) 
of the marginal objects are likely to be true globular clusters.
We therefore give a range of solutions for the completeness functions in 
Figure~\ref{fig-compfn},
corresponding to $f=0,0.5$ and 1.0. The figure shows, as expected, that existing catalogs
are reasonably complete to $V=18$, after which the completeness drops
drastically. To compute the completeness as a function of $R_{gc}$, we assumed that
detection efficiency at $R_{gc}>30$\arcmin\ was the same as that in the 
$R_{gc}=30$\arcmin\ bin. The completeness as a functions of $R_{gc}$ does not follow any 
particular pattern; the most important point is the low completeness in the innermost bin.
$C(R_{gc})$ can only be measured out to about $R_{gc}\le70$\arcmin,
and averaging over this region yields values for the overall
completeness of 50--85\%. The small number of objects per radial bin
and the uncertainty about the nature of the marginal objects make this
estimate rather imprecise.

It is important to know the total number of globular clusters in M31, since it
is one of the few spiral galaxies with well-studied GCSs. Existing surveys
\citep[summarized in][]{b00} and our new HST survey bring the number of confirmed
M31 GCs to over 250. The most comprehensive attempt to estimate the total
number of M31 GCs \citep{bat93} gives population ratios $N_{\rm M31}/N_{\rm MW}=2.5-3.5$;
with $N_{\rm MW}=150$, this gives $N_{\rm M31}=375-525$, or $450\pm75$.
We can use the results of our completeness study to attempt a new estimate
of the total number of M31 GCs. We take two approaches,
which use the completeness data somewhat differently.
One approach is to use the result that for $V<18$ and $R_{gc}>5$\arcmin, 
the existing sample is close to complete. We can therefore use the results of
GCLF fitting in this region to estimate the number of clusters fainter than $V=18$.
The GCLFs computed in \citet{bhb01} give a total number of clusters $N_{gc}$
in the range 394--417; the midpoint of the range is 406.
In the region $R\leq5$\arcmin, there are 37 cataloged 
clusters or candidates and the catalog completeness is about 70\%. This implies that the
true number of clusters is about 53, so the total number of GCs in
M31 is approximately $406+53=459$. A reasonable estimate of the error
in this value is 15\%, or $\pm69$.

We can also use the completeness estimates directly, to estimate 
\begin{equation}
N_{\rm gc} = \sum_{R_{gc}}\frac{N_{\rm good}+fN_{\rm marg}}{C(R_{gc})}
\end{equation}
where $N_{\rm good}$ and $N_{\rm marg}$ are the number of good 
and marginal cataloged clusters in a given $R_{gc}$ bin. The $C$ used
in the computation is the value plotted in Figure~\ref{fig-compfn}
for the appropriate value of $f$.
The catalog used is that given by \citet{b00} with likely NGC~205 clusters 
and (likely young) blue clusters removed.
The 299 good clusters are those confirmed by spectroscopy or 
high-resolution imaging and/or members of the `adopted best sample' 
of \citet{bat93}; the other 130 clusters are considered marginal.
The resulting $N_{gc}$ is sensitive to the value of $f$ and
ranges from $415\pm57$ for $f=0$ to $856\pm126$ for $f=1.0$. 
A value of $f=0.25$, which we believe is reasonable,
gives $N_{gc}=494\pm45$. Our two estimates of the total number of GCs
in M31 are compatible both with each other and with the results
of \citet{bat93}. The precision of our results is not much 
better than that of previous estimates, and improvement will 
require a wide-field, deep CCD survey of M31 which can be used to
find GCs in a uniform manner across the galaxy; such a survey
is currently being carried out \citep{lee01}.

We now consider implications of our estimated value of M31's $N_{gc}$
for its specific frequency $S_N$ and `mass specific frequency' $T$
\citep{za93}. To do this we need values for M31's luminosity and
mass-to-light ratio. \citet{ken87} gives the total magnitude of M31 as $V=3.28$.
Correcting for foreground extinction $A_V=0.25$ \citep[see][]{b00} 
and our adopted value of $(m-M)_0=24.47$ gives $M_V=-21.43$,
which is bracketed by the values given by \citet{vdblg}, $M_V=-21.2$,
and \citet{az98}, $M_V=-21.8$. For ease of comparison we use 
$M/L_V=6.1$ as do \citet{kp99}.
With $N_{gc}=459\pm69$, this gives $S_N=1.2\pm0.2$ and $T=2.4\pm0.4$. 
\citet{kp99} give $S_N$ and $T$ values for seven Sb-Sc spirals
in addition to M31. The mean and dispersion of $S_N$ and $T$ 
for these galaxies are
$\langle S_N\rangle=0.8\pm0.2$ and $\langle T\rangle=1.5\pm0.3$. 
For four Sa and Sab spirals the (highly uncertain) mean values are
$\langle S_N\rangle=2.0\pm0.6$ and $\langle T\rangle=4.0\pm1.1$. 
M31's values of $S_N$ and $T$ fall at the large end of 
the range of observed values for Sb and Sc spirals, and well within the
range observed for Sa and Sab spirals. While it has about twice as
many clusters per unit mass or luminosity than the Milky Way,
M31 is within the range of variation seen in other spirals' GCSs.
It is interesting to speculate on the difference between the Milky Way
in M31 in terms of differences in the two galaxies' histories, since
obviously environmental differences cannot be a major player.
\citet{fre99} suggested that perhaps M31 suffered an early major merger;
perhaps this was responsible for the creation of extra GCs in M31,
as in the picture of \citet{az92}. This would be consistent with the
suggestion that some of the metal-rich GCs in M31 are younger than
the rest of the population \citep{bh00,bhb01}, although there are not enough
metal-rich clusters to account for the entire `cluster excess' in M31.
Further explanation of the total number of clusters in M31 awaits
both a larger spiral comparison sample and more detailed theoretical
picture of GCS formation.

\section{Summary}

Using the Hubble Space Telescope Archive to search for M31 globular clusters
in WFPC2 images, we present the discovery of many previously-known clusters,
a number of new cluster candidates, and some 20 objects which may be M31 
open clusters. We use the discovery data, together with an estimate of
our discovery efficiency, to estimate the completeness of existing cluster catalogs.
As expected, the existing catalogs are least complete for faint
clusters and clusters very near the center of M31. As we found in 
a preliminary version of this analysis in \citet{bhb01}, the completeness is
very high to the magnitude limit we used for computing the
globular cluster luminosity function. This validates our finding that
the M31 GCLF varies with both radial distance from the center of
M31 and with metallicity.

We use the completeness results to estimate the total number of globular 
clusters in M31 and derive values in the range 450--500, consistent with
or somewhat higher than previous estimates. The specific frequency of
GCs in M31 is $S_N=1.2\pm0.2$ and the mass specific frequency $T=2.4\pm0.4$. 
M31 has more clusters per unit mass or luminosity than the Milky Way,
but is within the range of specific frequencies seen in the limited
number of other spirals studied to date.

\acknowledgments
We thank R. Di Stefano, J. Grindlay, D. Sasselov and S. Zepf 
for helpful discussions.

\clearpage

\clearpage

\begin{figure}
\caption
{HST images of M31 globular clusters. In row order, from top left:
006--058, 064--125, 077--138, 146--000, 156--211, 311--033, 331--057, 468--000, 000--001.
All images are in filter F555W or F606W except those of 064--125 and 146--000 (in F300W).
All images are 5\arcsec\ square; 077--138 is not centered in its image because
it fell near the edge of a WFPC2 chip.
\label{fig-gc}}
\end{figure}

\begin{figure}
\caption
{New globular cluster candidates found in HST images. All images
are 5\arcsec\ square. 
\label{fig-newclust}}
\end{figure}

\begin{figure}
\caption
{Marginal objects found in HST images: these objects are non-stellar
but not obviously star clusters. All images are 5\arcsec\ square. 
\label{fig-myst}}
\end{figure}

\begin{figure}
\caption
{Possible M31 open clusters found in HST images. All images
are 5\arcsec\ square. 
\label{fig-oc}}
\end{figure}

\clearpage

\begin{deluxetable}{lllll}
\tablewidth{0pt}
\tablenum{1}
\tablecaption{HST fields used in the search\label{tbl-flds}}
\tablehead{\colhead{RA(2000)}&\colhead{Dec (2000)}&\colhead{filter}&\colhead{Exposure\tablenotemark{a}}&\colhead{Dataset name}}
\startdata
00 32 36.21 &+39 27 43.4 &F606W&   1400 & U4K2OI01R\\	   
00 32 36.62 &+39 27 42.0 &F606W&   1500 & U4K2OI02R\\	   
00 32 49.01 &+39 35 00.4 &F555W&   1600 & U2E20709T\\	   
00 34 13.68 &+39 23 26.5 &F814W&   2800 & U2TA0501T\\	   
00 34 13.26 &+39 23 48.4 &F555W&   600  & U4490401R\\	   
00 34 13.46 &+39 24 40.5 &F702W&   600  & U27L0501T\\	   
00 36 59.20 &+39 52 21.3 &F555W&   800  & U4710201M\\	   
00 37 43.08 &+39 58 00.6 &F336W&   200  & U4F50907R\\	   
00 37 49.14 &+40 06 29.2 &F555W&   600  & U2782X01T\\	   
00 37 58.50 &+39 58 32.8 &F606W&   2100 & U67FFP01R\\	       
00 38 32.54 &+41 28 45.4 &F555W&   830  & U39I0104T\\	   
00 38 55.51 &+40 20 41.1 &F606W&   800  & U2804I01T\\	   
00 39 32.23 &+40 30 48.1 &F555W&   5300 & U4CA0701R\\	   
00 39 47.35 &+40 31 58.0 &F555W&   1200 & U5BJ0101R\\	   
00 39 53.99 &+41 47 19.2 &F555W&   2600 & U3KL1004R\\	   
00 40 01.58 &+40 34 14.8 &F555W&   1200 & U5BJ0201R\\	   
00 40 10.11 &+40 46 08.9 &F814W&   200  & U4WOAH05R\\	   
00 40 14.10 &+40 37 11.4 &F555W&   160  & U2YE0703T\\	   
00 40 14.86 &+40 49 02.8 &F814W&   200  & U4WOC605R\\	   
00 40 15.76 &+40 36 48.1 &F300W&   1200 & U2M80C01T\\	   
00 40 22.15 &+41 41 38.4 &F336W&   400  & U2GH020CT\\	   
00 40 23.16 &+41 40 55.6 &F555W&   2600 & U3KL0704M\\	   
00 40 23.66 &+41 41 55.2 &F555W&   2600 & U3KL0804R\\	   
00 40 23.77 &+41 41 40.8 &F555W&   100  & U2EE0506T\\	   
00 40 25.50 &+41 42 25.7 &F555W&   2600 & U3KL0904R\\	   
00 40 26.84 &+41 27 27.3 &F555W&   2000 & U2830103T\\	   
00 40 29.18 &+41 36 31.9 &F814W&   5400 & U3KL0501M\\	   
00 40 29.40 &+40 43 58.3 &F814W&   400  & U2AB0103T\\	   
00 40 30.61 &+40 44 50.5 &F336W&   400  & U4F51107R\\	   
00 40 31.26 &+40 42 59.6 &F336W&   400  & U4F51007R\\	   
00 40 33.17 &+40 45 39.0 &F606W&   350  & U2G20701T\\	   
00 40 33.81 &+41 39 40.2 &F555W&   2600 & U3KL0604M\\	   
00 40 34.22 &+41 22 06.5 &F555W&   5300 & U4CA0201R\\	   
00 40 39.54 &+40 33 25.5 &F555W&   520  & U34L6903R\\	   
00 40 39.75 &+40 53 24.0 &F555W&   300  & U2G20E03T\\	   
00 40 46.06 &+39 35 01.0 &F814W&   130  & U4WOAU05R\\	   
00 40 50.80 &+40 41 16.7 &F555W&   400  & U2Q00101T\\	   
00 40 56.68 &+40 35 29.0 &F555W&   5300 & U4CA0101R\\	   
00 40 59.08 &+40 46 42.1 &F606W&   1050 & U581OL01R\\	   
00 40 59.54 &+41 03 38.4 &F439W&   800  & U2TR0804T\\	   
00 41 16.28 &+40 56 12.6 &F555W&   5300 & U4CA0301R\\	   
00 41 17.85 &+41 09 00.7 &F814W&   3700 & U2OT0O01T\\	   
00 41 22.08 &+40 37 06.7 &F555W&   1200 & U5BJ0301R\\	   
00 41 38.85 &+39 35 39.8 &F814W&   130  & U4WOBK05R\\	   
00 41 42.21 &+40 12 22.4 &F814W&   2000 & U2830201T\\	   
00 41 43.30 &+41 34 20.4 &F555W&   2000 & U2830303T\\	   
00 41 53.85 &+40 50 30.2 &F814W&   1200 & U2806A02T\\	   
00 41 55.58 &+40 47 15.0 &F555W&   5300 & U4CA0601R\\	   
00 42 05.02 &+41 12 14.9 &F300W&   2300 & U2OU7501T\\	   
00 42 05.27 &+40 57 33.9 &F555W&   520  & U34L7003R\\	   
00 42 06.07 &+41 07 55.5 &F814W&   6200 & U3B83Y01T\\	   
00 42 14.14 &+41 10 22.6 &F606W&   160  & U2OURQ01T\\	   
00 42 14.36 &+41 06 24.7 &F606W&   1800 & U581R201R\\	   
00 42 14.40 &+41 10 11.7 &F555W&   8400 & U3D90207T\\	   
00 42 18.01 &+40 45 03.7 &F555W&   900  & U3DG0107T\\	   
00 42 27.21 &+41 08 28.0 &F300W&   1600 & U2OUUT01T\\	   
00 42 28.88 &+41 03 05.2 &F606W&   800  & U4K2RG01R\\	   
00 42 31.00 &+41 10 12.2 &F814W&   400  & U4WO9N05R\\	   
00 42 32.47 &+41 13 39.5 &F555W&   5200 & U2Y30204T\\	   
00 42 32.70 &+40 33 55.5 &F555W&   1200 & U3YK0101R\\	   
00 42 35.13 &+41 10 35.1 &F555W&   5200 & U2Y30305T\\	   
00 42 38.97 &+41 15 29.2 &F555W&   1680 & U2KJ0109T\\	   
00 42 39.28 &+40 51 42.2 &F814W&   600  & U2E20401T\\	   
00 42 39.49 &+40 51 46.9 &F555W&   104  & U2LG0101T\\	   
00 42 39.88 &+41 10 48.9 &F814W&   400  & U42Z2302R\\	   
00 42 40.85 &+41 15 51.2 &F555W&   2500 & U5LT0104R\\	   
00 42 40.96 &+40 51 07.3 &F555W&   110  & U2EE0405T\\	   
00 42 41.68 &+40 51 04.3 &F555W&   2000 & U2880704T\\	   
00 42 41.74 &+41 15 57.9 &F814W&   600  & U2E20201T\\	   
00 42 42.21 &+40 52 22.4 &F555W&   1200 & U2E20307T\\	   
00 42 44.64 &+41 16 39.2 &F555W&   1680 & U2E2010BT\\	   
00 42 46.91 &+41 16 15.9 &F555W&   2200 & U2LG0201T\\	   
00 42 47.63 &+41 16 11.0 &F336W&   460  & U2LH0103T\\	   
00 42 50.04 &+41 36 17.8 &F814W&   700  & U4WOC805R\\	   
00 42 50.15 &+40 59 56.0 &F814W&   8000 & U2OQ3201T\\	   
00 42 50.34 &+41 17 54.4 &F814W&   7000 & U2OQF801T\\	   
00 42 51.45 &+41 06 52.9 &F814W&   400  & U4WOA305R\\	   
00 42 52.07 &+41 24 53.4 &F814W&   6700 & U2OQF301T\\	   
00 42 52.26 &+41 08 06.8 &F814W&   4500 & U26KCZ01T\\	   
00 42 52.37 &+41 10 31.7 &F814W&   400  & U42Z3402R\\	   
00 42 52.73 &+40 56 30.4 &F814W&   2000 & U4XI0101R\\	   
00 42 53.03 &+41 14 23.4 &F814W&   400  & U42Z1102R\\	   
00 42 54.11 &+41 08 08.9 &F814W&   800  & U4WOA205R\\	   
00 42 54.87 &+41 10 35.0 &F814W&   700  & U4WOBC05R\\	   
00 42 58.84 &+40 50 34.3 &F814W&   4400 & U3VJ0103R\\	   
00 43 00.93 &+41 13 17.7 &F300W&   2600 & U31K0109T\\	   
00 43 01.94 &+41 19 19.9 &F555W&   5200 & U38K0103T\\	   
00 43 04.61 &+40 54 33.0 &F555W&   2000 & U2880801T\\	   
00 43 05.28 &+40 50 37.8 &F300W&   5400 & U27H0F01T\\	   
00 43 05.38 &+40 56 40.4 &F606W&   350  & U4G40104R\\	   
00 43 06.12 &+41 12 59.7 &F555W&   2000 & U5850103R\\	   
00 43 06.14 &+41 13 00.0 &F555W&   2000 & U5850107R\\	   
00 43 06.19 &+40 56 52.0 &F606W&   350  & U4G40103R\\	   
00 43 06.40 &+40 56 31.2 &F814W&   350  & U4G40101R\\	   
00 43 07.21 &+40 56 42.8 &F814W&   260  & U4G40102R\\	   
00 43 07.85 &+40 53 32.8 &F814W&   1100 & U4C80403R\\	   
00 43 08.59 &+41 14 51.7 &F300W&   8400 & U27H0E01T\\	   
00 43 09.06 &+40 51 18.6 &F814W&   1000 & U4WOBJ05R\\	   
00 43 12.49 &+41 02 02.7 &F606W&   350  & U4G40204R\\	   
00 43 12.82 &+41 02 17.0 &F606W&   350  & U4G40203R\\	   
00 43 13.76 &+41 01 59.0 &F814W&   350  & U4G40201R\\	   
00 43 14.09 &+41 02 13.3 &F814W&   260  & U4G40202R\\	   
00 43 18.05 &+39 49 13.1 &F814W&   2000 & U2830401T\\	   
00 43 20.30 &+41 05 36.2 &F814W&   400  & U42Z1202R\\	   
00 43 20.80 &+41 06 14.5 &F814W&   300  & U42Z4602R\\	   
00 43 22.39 &+41 13 53.8 &F814W&   1500 & U2OT0S01T\\	   
00 43 25.28 &+41 04 02.2 &F814W&   4400 & U3VJ0203R\\	   
00 43 36.72 &+41 26 15.4 &F555W&   350  & U2KW0601T\\	   
00 43 43.26 &+41 00 32.4 &F814W&   400  & U4WOA105R\\	   
00 43 46.58 &+41 11 14.7 &F814W&   300  & U42Z5702R\\	   
00 43 47.85 &+41 11 00.6 &F814W&   300  & U42Z5802R\\	   
00 43 54.55 &+41 24 10.8 &F300W&   600  & U2M80G01T\\	   
00 43 57.06 &+41 25 33.4 &F300W&   600  & U2M80H01T\\	   
00 44 14.38 &+41 20 45.2 &F336W&   400  & U4F51207R\\	   
00 44 23.45 &+41 20 40.8 &F336W&   320  & U4F51307R\\	   
00 44 23.74 &+41 45 16.3 &F555W&   800  & U4710101M\\	   
00 44 34.81 &+41 38 38.4 &F336W&   800  & U5750101R\\	   
00 44 35.25 &+41 31 21.6 &F814W&   500  & U4WOBH08R\\	   
00 44 36.29 &+41 35 05.6 &F814W&   500  & U4WOBS05R\\	   
00 44 42.45 &+41 44 24.2 &F555W&   5300 & U4CA0401M\\	   
00 44 42.52 &+41 44 24.1 &F555W&   5300 & U4CA0501R\\	   
00 44 42.59 &+41 14 30.3 &F814W&   500  & U4WOA005R\\	   
00 44 44.23 &+41 27 33.9 &F555W&   140  & U2Y20106T\\	   
00 44 46.19 &+41 51 33.4 &F555W&   1200 & U5BJ0401R\\	   
00 44 49.34 &+41 28 59.0 &F555W&   140  & U2Y20206T\\	   
00 44 50.61 &+41 19 11.1 &F555W&   3800 & U2GV0401T\\	   
00 44 51.22 &+41 30 03.7 &F555W&   160  & U2YE0603T\\	   
00 44 57.63 &+41 30 51.7 &F555W&   140  & U2Y20306T\\	   
00 45 00.36 &+41 31 55.1 &F300W&   600  & U2M80A01T\\	   
00 45 03.79 &+41 31 09.6 &F300W&   600  & U2M80E01T\\	   
00 45 05.66 &+41 38 00.4 &F336W&   320  & U4F51407R\\	   
00 45 07.76 &+41 36 46.8 &F336W&   280  & U4F51507R\\	   
00 45 09.25 &+41 34 30.7 &F555W&   140  & U2Y20406T\\	   
00 45 11.95 &+41 36 57.0 &F555W&   140  & U2Y20506T\\	   
00 45 28.46 &+41 05 53.9 &F555W&   4320 & U2UG010AT\\	   
00 45 36.98 &+41 42 17.3 &F606W&   2300 & U5HNM301R\\	    
00 45 38.15 &+41 36 02.4 &F300W&   650  & U4WOBG0ER\\	   
00 45 39.00 &+41 36 36.3 &F814W&   580  & U4WOBG09R\\	   
00 45 39.25 &+41 36 32.4 &F606W&   1140 & U4WOBG01R\\	   
00 46 01.64 &+40 41 58.3 &F555W&   4320 & U2UG020AT\\	   
00 46 20.46 &+40 16 34.1 &F555W&   5300 & U4CA0801R\\	   
00 46 24.33 &+42 07 01.7 &F814W&   350  & U4WOC905R\\	   
00 46 24.56 &+42 01 38.7 &F555W&   5300 & U4CA0901R\\	   
00 46 29.85 &+42 04 50.4 &F336W&   320  & U4F51707R\\	   
00 48 15.15 &+40 26 31.0 &F606W&   1000 & U36Z7801R\\	   
00 48 21.29 &+40 29 02.4 &F606W&   1000 & U36Z8401R\\	   
00 48 42.83 &+40 24 47.2 &F606W&   1000 & U36Z7701R\\	   
00 48 44.22 &+40 33 06.0 &F606W&   1000 & U36Z8201R\\	   
00 49 08.22 &+40 30 33.9 &F606W&   1000 & U36Z8301R\\	   
00 49 13.48 &+40 21 51.1 &F606W&   1000 & U36Z8001R\\	   
00 49 18.35 &+40 30 12.0 &F606W&   1000 & U36Z8501R\\	   
00 49 28.48 &+40 21 54.4 &F606W&   1000 & U36Z8101R\\	   
00 49 31.26 &+40 27 57.9 &F606W&   1000 & U36Z8601R\\	   
00 50 06.13 &+41 33 56.0 &F606W&   2100 & U67FGY01R
\enddata
\tablenotetext{a}{`Exposure' is the combined exposure time, in seconds, of 
all images in the specified filter and position.}
\end{deluxetable}

\begin{deluxetable}{lll}
\tablewidth{0pt}
\tablenum{2}
\tablecaption{Cataloged objects which are not M31 globular clusters\label{tbl-noncl}}
\tablehead{\colhead{name}&\colhead{class\tablenotemark{a}}&\colhead{HST field(s)}}
\startdata
000-253   &star       & U2M80H01T[4]\\    
000-M046  &HII region?& U2Y20106T[2]\\ 
000-M050  &HII region?& U4WOBH08A[4]\\ 
000-M068  &stars      & U4F51407A[4]\\ 
000-V211  &stars      & U2TR0804B[4]\\ 
000-V212  &star       & U2TR0804B[1]\\ 
000-V298  &star       & U2Y20206T[4]\\ 
064D-NB80 &star       & U2KJ0109A[3]\\ 
074D-NB88 &star       & U2LH0103B[4] U2LG0201B[4] U42Z1102R[3]\\
075D-NB96 &star       & U2LH0103B[4] U2LG0201B[4]\\ 
080D-NB93 &star       & U2LH0103B[3] U2LG0201B[3]\\ 
084D-000  &star       & U2OQF801A[3]\\ 
086D-000  &star       & U3VJ0103A[3]\\ 
093D-000  &star       & U27H0F01B[3]\\ 
114D-000  &star       & U4710101A[4]\\ 
133-191   &stars      & U2OQF301A[4]\\ 
138-000   &star       & U2OQF801A[3]\\ 
166-000   &star       & U2OT0S01A[4]\\ 
185D-000  &star       & U5BJ0201A[3]\\ 
193D-055  &star       & U2EE0506T[3] U3KL0804A[4] U3KL0904A[1]\\   
254D-000  &star       & U5BJ0401A[4]\\ 	 	    
285-000   &galaxy     & U2UG010AA[2]\\ 		 
326-000   &star       & U2EE0506T[1] U3KL0804A[2] U2GH020CA[1]\\
332-000   &star       & U3KL0704A[4]\\ 		 
444-000   &star       & U2EE0506T[2] U3KL0704A[2] U3KL0804A[3]\\   
446-000   &galaxy     & U3KL0501A[4]\\ 
501-345   &galaxy     & U36Z8401A[1]\\ 
NB100     &star       & U31K0109B[3]\\ 
NB103  	  &star       & U42Z1102R[4]\\ 
NB104 	  &star       & U42Z1102R[1]\\ 
NB106 	  &star       & U31K0109B[4]\\ 
NB26      &star       & U31K0109B[3]\\ 
NB27 	  &star (pr)  & U2Y30204A[2]\\ 
NB30 	  &star (pr)  & U2E2010BA[3]\\ 
NB37 	  &star       & U2KJ0109A[4]\\ 
NB42      &star (pr)  & U2E2010BA[3] U2LH0103B[2] U2OQF801A[1]\\ 
NB44      &star       & U2E2010BA[3]\\ 
NB45 	  &star       & U42Z1102R[3]\\ 
NB49 	  &star       & U42Z1102R[2] U31K0109B[3]\\
NB51 	  &stars (pr) & U2Y30204A[3]\\ 
NB53 	  &star       & U2KJ0109A[3]\\ 
NB54 	  &star       & U2Y30204A[3]\\ 
NB56 	  &star       & U2Y30204A[4]\\ 
NB75      &star       & U2KJ0109A[4]\\ 
NB76      &star       & U2KJ0109A[4]\\ 
NB78      &star       & U2KJ0109A[3] U5LT0104A[3]\\
NB82      &stars (pr) & U2Y30204A[3]\\ 	
NB84      &star (pr)  & U2Y30204A[3]\\ 	
NB85      &star (pr)  & U2Y30204A[3]\\ 	
NB94 	  &star       & U2LH0103B[4] U2LG0201B[4]\\
NB95      &star       & U2LH0103B[4] U2LG0201B[4]\\
NB97 	  &star       & U42Z1102R[1]\\ 	
NB99 	  &star       & U31K0109B[3] U42Z1102R[2]\\
000-D040  &blank    & U4CA0101A[4]\\ 
000-M023  &blank    & U4F51307B[4]\\
000-M056  &blank    & U4F51407A[3]\\ 
000-M060  &blank    & U4F51407A[4]\\ 
038D-000  &blank    & U2OU7501T[3]\\
092D-000  &blank    & U31K0109B[2]\\
353-000   &blank    & U2LH0103B[4] U2LG0201B[4]  U5LT0104A[2]\\
NB18      &blank    & U2OQF801A[2]\\		  		 
NB38 	  &blank    & U2Y30204A[2]\\		  		 
NB40 	  &blank    & U42Z1102R[2]\\		  		 
NB57 	  &blank    & U2Y30204A[2]\\		  		 
NB58 	  &blank    & U5LT0104A[2] U2E20201A[2]  U2KJ0109A[2]\\
NB59 	  &blank    & U2LH0103B[4] U42Z1102R[4] \\ 		 
NB74      &blank    & U2E2010BA[2] U2E20201A[4]  U5LT0104A[4]\\
NB87      &blank    & U2OQF801A[2]\\
\enddata
\tablenotetext{a}{`pr' refers to objects which have slightly larger FWHMs than
most stars, although we still believe them to be stars.
`stars' refers to objects which appear to be blended images of 2 or 3 stars.
`Blank' refers to a object which was not detected at its catalog coordinates.}
\end{deluxetable}

\begin{deluxetable}{lllll}
\tablewidth{0pt}
\tablenum{3}
\tablecaption{New globular cluster candidates found in M31 HST fields\label{tbl-newgc}}
\tablehead{\colhead{name}&\colhead{RA(2000)}&\colhead{Dec (2000)}&\colhead{quality}&\colhead{comments}}
\startdata
M31GC~J003411+392359& 00 34 11.48&39 23 59.1 &C/D&\nodata\\
M31GC~J004010+403625& 00 40 10.33&40 36 24.7 &C/D&\nodata\\
M31GC~J004023+414045& 00 40 22.68&41 40 44.5 &C  &\nodata\\
M31GC~J004027+414225& 00 40 27.25&41 42 24.8 &B  &\nodata\\
M31GC~J004030+404530& 00 40 30.46&40 45 29.6 &B  &\nodata\\
M31GC~J004031+404454& 00 40 30.63&40 44 54.3 &C  &\nodata\\
M31GC~J004031+412627& 00 40 30.68&41 26 27.1 &C  &\nodata\\
M31GC~J004034+413905& 00 40 34.42&41 39 04.8 &C/D&\nodata\\
M31GC~J004037+403321& 00 40 37.15&40 33 21.4 &C  &\nodata\\
M31GC~J004045+405308& 00 40 44.92&40 53 07.6 &C  &\nodata\\
M31GC~J004051+404039& 00 40 50.68&40 40 38.6 &B/C&\nodata\\
M31GC~J004103+403458& 00 41 02.88&40 34 57.9 &B  &Hodge 119?\\
M31GC~J004146+413326& 00 41 45.57&41 33 26.2 &C  &\nodata\\
M31GC~J004200+404746& 00 42 00.39&40 47 45.8 &C  &\nodata\\
M31GC~J004228+403330& 00 42 27.56&40 33 29.8 &C/D&\nodata\\
M31GC~J004246+411737& 00 42 46.01&41 17 36.5 &C  &\nodata\\
M31GC~J004251+405841& 00 42 50.80&40 58 40.7 &C  &\nodata\\
M31GC~J004251+411035& 00 42 50.78&41 10 34.7 &A  &\nodata\\
M31GC~J004257+404916& 00 42 57.05&40 49 16.4 &C  &Hodge 195?\\
M31GC~J004258+405645& 00 42 58.02&40 56 45.4 &A  &\nodata\\
M31GC~J004301+405418& 00 43 01.35&40 54 17.5 &B  &\nodata\\
M31GC~J004304+405129& 00 43 04.27&40 51 29.2 &C  &\nodata\\
M31GC~J004304+412028& 00 43 03.75&41 20 28.2 &A  &\nodata\\
M31GC~J004312+405303& 00 43 11.86&40 53 02.8 &B  &\nodata\\
M31GC~J004312+410249& 00 43 11.99&41 02 49.1 &C  &\nodata\\
M31GC~J004424+414502& 00 44 23.71&41 45 02.3 &C  &X-ray src: SHP278?\\
M31GC~J004425+414529& 00 44 25.21&41 45 29.1 &C/D&\nodata\\
M31GC~J004439+414426& 00 44 39.07&41 44 26.3 &C  &\nodata\\
M31GC~J004537+413644& 00 45 37.25&41 36 44.3 &B  &\nodata\\
M31GC~J004537+414332& 00 45 36.75&41 43 32.2 &C  &\nodata\\
M31GC~J004622+420631& 00 46 21.80&42 06 30.8 &C  &\nodata\\
M31GC~J004624+420059& 00 46 23.50&42 00 58.5 &C  &\nodata\\ 
M31OC~J003836+412739& 00 38 35.73&41 27 39.3 &B  &\nodata\\ 
M31OC~J003941+403154& 00 39 40.52&40 31 53.6 &C  &\nodata\\ 
M31OC~J003943+403116& 00 39 43.21&40 31 15.6 &C  &\nodata\\ 
M31OC~J004000+403326& 00 39 59.99&40 33 25.9 &C  &\nodata\\ 
M31OC~J004008+403507& 00 40 07.55&40 35 06.6 &B  &\nodata\\ 
M31OC~J004027+404524& 00 40 27.26&40 45 23.7 &C  &\nodata\\ 
M31OC~J004031+404537& 00 40 30.51&40 45 37.4 &C  &\nodata\\ 
M31OC~J004053+403519& 00 40 52.94&40 35 19.2 &D  &\nodata\\ 
M31OC~J004054+404625& 00 40 54.14&40 46 24.7 &C  &\nodata\\ 
M31OC~J004057+403425& 00 40 56.62&40 34 24.7 &C  &\nodata\\ 
M31OC~J004119+403608& 00 41 18.69&40 36 08.2 &B/C&\nodata\\ 
M31OC~J004123+403756& 00 41 23.30&40 37 56.1 &C  &\nodata\\ 
M31OC~J004421+414516& 00 44 21.44&41 45 15.9 &C  &\nodata\\ 
M31OC~J004442+415122& 00 44 41.84&41 51 22.4 &C  &\nodata\\ 
M31OC~J004442+415237& 00 44 42.25&41 52 36.7 &C  &\nodata\\ 
M31OC~J004449+414430& 00 44 48.83&41 44 30.3 &C/D&\nodata\\ 
M31OC~J004450+415211& 00 44 50.27&41 52 11.1 &C  &\nodata\\ 
M31OC~J004510+413646& 00 45 10.45&41 36 46.3 &C  &Hodge 311?\\ 
M31OC~J004512+413712& 00 45 11.81&41 37 11.6 &C  &H~II region, Hodge 313?\\ 
M31OC~J004539+414220& 00 45 38.88&41 42 20.4 &C  &Radio src MY0042+414?\\ 
\enddata
\end{deluxetable}

\begin{deluxetable}{lllllll}
\tablewidth{0pt}
\tablenum{4}
\tablecolumns{7}
\tablecaption{Photometry of new clusters and candidates in M31 HST fields\label{tbl-phot}}
\tablehead{\colhead{name}&\colhead{$U$\tablenotemark{a}}&\colhead{$B$}
&\colhead{$V$}&\colhead{$R$}&\colhead{$I$}&\colhead{$\langle r_{1/2}\rangle$ (\arcsec)}}
\startdata
\sidehead{Cataloged clusters}
000--001  &\nodata &\nodata &13.807   &\nodata&12.684  &0.40\\
000--D38  &\nodata &\nodata &19.247   &\nodata&18.276  &0.37\\
000--M045 &\nodata &19.391  &18.723   &\nodata&17.446  &1.49\\
000--M91  &\nodata &\nodata &19.143   &\nodata&\nodata &0.89\\
006--058  &\nodata &\nodata &15.463   &\nodata&14.354  &0.38\\
009--061  &\nodata &\nodata &\nodata  &\nodata&15.809  &0.71\\
011--063  &\nodata &\nodata &16.578   &\nodata&15.624  &0.31\\
012--064  &\nodata &\nodata &15.042   &\nodata&13.979  &0.42\\
018--071  &\nodata &\nodata &17.533   &\nodata&16.385  &1.14\\
020D--089 &\nodata &\nodata &\nodata  &\nodata&16.039  &0.80\\
027--087  &\nodata &\nodata &15.559   &\nodata&14.409  &0.41\\
030--091  &\nodata &\nodata &17.377   &\nodata&15.593  &0.59\\
045--108  &\nodata &\nodata &15.784   &\nodata&14.477  &0.42\\
058--119  &\nodata &\nodata &14.925   &\nodata&13.837  &0.36\\
064--125  &17.461* &\nodata &\nodata  &\nodata&\nodata &0.69\\
068--130  &\nodata &17.575  &16.407   &\nodata&14.849  &0.68\\
070--133  &\nodata &\nodata &17.262   &\nodata&16.165  &0.20\\
071--000  &22.716* &\nodata &\nodata  &\nodata&\nodata &0.07\\
076--138  &17.720  &17.483  &16.927   &\nodata&15.626  &0.53\\
077--139  &\nodata &18.829  &17.734   &\nodata&16.125  &0.42\\
092--152  &18.766* &\nodata &\nodata  &\nodata&\nodata &0.41\\
097D--000 &\nodata &\nodata &\nodata  &17.878 &17.121  &1.08\\
101--164  &18.523* &\nodata &\nodata  &\nodata&\nodata &0.42\\
109--170  &\nodata &17.407  &16.197   &\nodata&14.936  &0.61\\
110--172  &\nodata &\nodata &15.355   &\nodata&\nodata &0.66\\
114--175  &\nodata &\nodata &17.439   &\nodata&15.940  &0.47\\
115--177  &\nodata &\nodata &15.997   &\nodata&14.560  &0.25\\
118--NB6  &\nodata &\nodata &16.431   &\nodata&15.207  &0.52\\
123--182  &\nodata &\nodata &17.416   &16.754 &16.126  &0.58\\
124--NB10 &16.094  &\nodata &14.777   &\nodata&13.631  &0.53\\
127--185  &15.756  &\nodata &14.467   &\nodata&13.239  &0.75\\
128--187  &\nodata &\nodata &\nodata  &16.441 &15.764  &0.41\\
132--000  &\nodata &\nodata &17.739   &17.244 &16.440  &0.34\\
134--190  &\nodata &\nodata &\nodata  &16.064 &15.502  &0.52\\
143--198  &\nodata &\nodata &15.954   &\nodata&14.731  &0.25\\
145--000  &19.901* &\nodata &\nodata  &\nodata&\nodata &0.47\\
146--000  &18.458* &\nodata &\nodata  &\nodata&\nodata &0.59\\
148--200  &\nodata &\nodata &16.110   &\nodata&\nodata &0.51\\
153--000  &18.220* &\nodata &\nodata  &\nodata&\nodata &0.38\\
155--210  &\nodata &\nodata &18.011   &\nodata&16.672  &0.40\\
156--211  &\nodata &\nodata &16.969   &\nodata&15.873  &0.64\\
160--214  &\nodata &\nodata &18.076   &\nodata&17.075  &0.52\\
167--000  &\nodata &\nodata &\nodata  &\nodata&16.109  &0.33\\
205--256  &16.938* &\nodata &\nodata  &\nodata&\nodata &0.32\\
231--285  &\nodata &18.227  &17.248   &\nodata&\nodata &0.48\\
232--286  &16.440  &16.391  &15.646   &\nodata&14.543  &0.66\\
233--287  &\nodata &\nodata &15.718   &\nodata&14.585  &0.41\\
234--290  &\nodata &17.780  &16.783   &\nodata&\nodata &0.51\\
240--302  &\nodata &\nodata &15.181   &\nodata&14.230  &0.80\\
257--000  &\nodata &11.907  &20.960   &\nodata&16.312  &0.65\\
264--000  &18.652* &\nodata &17.577   &\nodata&16.811  &0.67\\
268--000  &\nodata &\nodata &18.314   &\nodata&16.880  &0.39\\
279--D068 &\nodata &\nodata &18.549   &\nodata&16.964  &0.68\\
311--033  &\nodata &\nodata &15.445   &\nodata&14.215  &0.38\\
315--038  &\nodata &16.548  &16.473   &\nodata&\nodata &0.56\\
317--041  &\nodata &\nodata &16.573   &\nodata&15.713  &0.73\\
318--042  &\nodata &17.234  &17.047   &\nodata&\nodata &0.63\\
319--044  &\nodata &18.333  &17.608   &\nodata&\nodata &0.41\\
324--051  &\nodata &\nodata &18.446   &\nodata&17.633  &0.30\\
328--054  &\nodata &\nodata &17.861   &\nodata&16.918  &0.85\\
330--056  &\nodata &\nodata &17.724   &\nodata&16.555  &0.56\\
331--057  &\nodata &\nodata &18.191   &\nodata&17.076  &0.56\\
333--000  &\nodata &\nodata &18.840   &\nodata&17.711  &0.90\\
338--076  &\nodata &\nodata &14.195   &\nodata&13.174  &0.56\\
342--094  &\nodata &18.033  &17.733   &\nodata&\nodata &0.92\\
343--105  &\nodata &\nodata &16.310   &\nodata&15.274  &0.36\\
358--219  &\nodata &\nodata &15.219   &\nodata&14.122  &0.55\\
368--293  &\nodata &18.189  &17.924   &\nodata&\nodata &0.54\\
374--306  &19.128* &\nodata &18.319   &\nodata&\nodata &0.68\\
379--312  &\nodata &\nodata &16.183   &\nodata&14.936  &0.65\\
384--319  &\nodata &\nodata &15.752   &\nodata&14.564  &0.36\\
386--322  &\nodata &\nodata &15.547   &\nodata&14.393  &0.36\\
468--000  &\nodata &\nodata &17.788   &\nodata&16.626  &1.95\\
NB21      &\nodata &\nodata &17.865   &\nodata&16.771  &0.41\\
NB39      &18.551  &\nodata &17.941   &\nodata&17.876  &0.38\\
NB41   	  &\nodata &\nodata &18.097   &\nodata&17.183  &0.43\\
NB81      &\nodata &\nodata &17.025   &\nodata&\nodata &0.35\\
NB83   	  &\nodata &\nodata &17.585   &\nodata&16.599  &0.25\\
NB86   	  &\nodata &\nodata &18.522   &\nodata&17.446  &0.17\\
NB89      &\nodata &\nodata &17.965   &\nodata&16.888  &0.34\\
\sidehead{New clusters}
M31GC~J003411+392359  &\nodata &\nodata &\nodata  &22.302 & \nodata&4.44\\
M31GC~J004010+403625  &\nodata &18.906  &18.505   &\nodata&\nodata &0.24\\
M31GC~J004023+414045  &\nodata &\nodata &18.289   &\nodata&16.990  &1.08\\ 
M31GC~J004027+414225  &\nodata &\nodata &19.691   &\nodata&19.138  &0.76\\
M31GC~J004030+404530  &\nodata &\nodata &16.064   &\nodata&\nodata &0.37\\
M31GC~J004031+404454  &\nodata &15.090  &22.708   &\nodata&20.337  &0.29\\
M31GC~J004031+412627  &\nodata &\nodata &20.930   &\nodata&19.477  &0.57\\
M31GC~J004034+413905  &\nodata &\nodata &18.666   &\nodata&\nodata &7.52\\  
M31GC~J004037+403321  &\nodata &\nodata &19.773   &\nodata&18.773  &0.49\\
M31GC~J004051+404039  &\nodata &\nodata &19.862   &\nodata&18.616  &0.73\\
M31GC~J004103+403458  &\nodata &\nodata &18.487   &\nodata&17.920  &0.38\\
M31GC~J004146+413326  &\nodata &\nodata &20.716   &\nodata&18.826  &0.39\\
M31GC~J004200+404746  &\nodata &\nodata &20.327   &\nodata&19.636  &0.45\\
M31GC~J004228+403330  &\nodata &\nodata &21.195   &\nodata&19.477  &0.51\\
M31GC~J004246+411737  &19.507* &\nodata &18.111   &\nodata&17.706  &0.43\\
M31GC~J004251+405841  &\nodata &\nodata &20.296   &\nodata&19.309  &0.20\\
M31GC~J004251+411035  &\nodata &19.177  &18.178   &\nodata&16.886  &0.21\\
M31GC~J004257+404916  &\nodata &\nodata &20.271   &\nodata&18.648  &1.18\\
M31GC~J004258+405645  &\nodata &\nodata &18.066   &\nodata&\nodata &0.31\\
M31GC~J004301+405418  &\nodata &\nodata &18.613   &\nodata&17.290  &0.45\\
M31GC~J004304+405129  &\nodata &\nodata &19.666   &\nodata&18.392  &1.12\\
M31GC~J004304+412028  &\nodata &\nodata &18.828   &\nodata&16.857  &0.38\\
M31GC~J004312+405303  &\nodata &\nodata &20.670   &\nodata&19.150  &0.73\\
M31GC~J004312+410249  &\nodata &\nodata &18.810   &\nodata&18.816  &0.52\\
M31GC~J004424+414502  &\nodata &\nodata &21.143   &\nodata&19.232  &0.31\\
M31GC~J004425+414529  &\nodata &\nodata &20.597   &\nodata&20.580  &0.99\\
M31GC~J004439+414426  &\nodata &\nodata &19.840   &\nodata&18.746  &0.67\\
M31GC~J004537+413644  &\nodata &20.350  &19.648   &\nodata&18.724  &0.92\\
M31GC~J004537+414332  &22.529* &\nodata &19.759   &\nodata&\nodata &0.47\\
M31GC~J004622+420631  &\nodata &20.085  &18.614   &\nodata&17.135  &2.86\\
M31GC~J004624+420059  &\nodata &\nodata &20.445   &\nodata&19.188  &0.70\\
M31OC~J004539+414220  &20.812* &\nodata &20.163   &\nodata&\nodata &0.61\\
M31OC~J004027+404524  &\nodata &\nodata &18.148   &\nodata&18.046  &0.55\\
M31OC~J004512+413712  &\nodata &18.352  &17.466   &\nodata&\nodata &0.65\\
M31OC~J003941+403154  &\nodata &20.398  &20.002   &\nodata&\nodata &0.68\\
M31OC~J004123+403756  &\nodata &21.752  &19.965   &\nodata&\nodata &0.78\\
M31OC~J004442+415122  &\nodata &19.788  &19.533   &\nodata&\nodata &1.08\\
M31OC~J004442+415237  &\nodata &19.997  &19.812   &\nodata&\nodata &0.50\\
M31OC~J004450+415211  &\nodata &20.599  &20.439   &\nodata&\nodata &0.40\\
M31OC~J004449+414430  &\nodata &\nodata &20.229   &\nodata&19.455  &0.75\\
M31OC~J004054+404625  &\nodata &23.432  &22.082   &\nodata&21.504  &0.18\\
M31OC~J003943+403116  &\nodata &20.854  &21.014   &\nodata&\nodata &0.45\\
M31OC~J004000+403326  &\nodata &19.232  &19.006   &\nodata&\nodata &1.09\\
M31OC~J004031+404537  &\nodata &\nodata &17.806   &\nodata&\nodata &4.75\\
M31OC~J004057+403425  &\nodata &\nodata &18.756   &\nodata&18.675  &0.86\\
M31OC~J004053+403519  &\nodata &\nodata &18.579   &\nodata&18.634  &0.99\\
M31OC~J003836+412739  &\nodata &\nodata &20.406   &\nodata&\nodata &1.43\\
M31OC~J004008+403507  &\nodata &20.773  &20.298   &\nodata&\nodata &0.84\\
M31OC~J004119+403608  &\nodata &\nodata &19.501   &\nodata&\nodata &1.88\\
\enddata
\tablenotetext{a}{Asterisks indicate F300W, instead of standard $U$-band magnitudes.}
\end{deluxetable}

\end{document}